\documentclass{eptcs}
\providecommand{\event}{SSV 2012} % Name of the event you are submitting to

\usepackage{paralist}
\usepackage[dvipsnames]{xcolor}
\usepackage[color]{changebar}
\usepackage{comment}
\usepackage{keylogo}
\usepackage{subfig}
\usepackage{listings}
\usepackage[T1]{fontenc}
\usepackage[scaled]{beramono}
\usepackage{etoolbox}

\newcommand{\subpara}[1]{\smallskip\\\noindent{\em #1}.\ \ }

\makeatletter
\renewcommand{\paragraph}{%
  \@startsection{paragraph}{4}%
  {\z@}{.5ex \@plus .2ex \@minus .2ex}{-1em}%
  {\normalfont\normalsize\bfseries}%
}

\renewcommand{\subparagraph}{%
  \@startsection{subparagraph}{5}%
  {\parindent}{.5ex \@plus .2ex \@minus .2ex}{-1em}
  {\normalfont \normalsize \itshape}%
}
\makeatother

\newcommand{\VerisoftXT}{Verisoft~XT}

\title{Lessons Learned From Microkernel Verification\thanks{Work partially 
funded by the German Federal Ministry of Education and Research (BMBF) in the 
framework of the \VerisoftXT\ project under grant 01 IS 07 008. The 
responsibility for this article lies with the authors.}\\
{\Large Specification is the New Bottleneck}}

\author{
Christoph Baumann$^1$\quad  Bernhard Beckert$^2$\quad Holger Blasum$^3$\quad 
Thorsten Bormer$^2$\\[10pt]
$^1$Saarland University, Saarbr\"{u}cken, Germany\\
$^2$Karlsruhe Institute of Technology, Karlsruhe, Germany\\
$^3$SYSGO AG, Klein-Winternheim, Germany
}

\def\titlerunning{Lessons Learned From Microkernel Verification}
\def\authorrunning{C. Baumann, B. Beckert, H. Blasum \& T. Bormer}

\begin{document}
\maketitle

\begin{abstract}
Software verification tools have become a lot more powerful in recent
years. Even verification of large, complex systems is feasible, as 
demonstrated in the L4.verified and \VerisoftXT{} projects. Still, 
functional verification of large software systems is rare -- for reasons beyond the
large scale of verification effort needed due to the size alone. 
In this paper we report on lessons learned for verification of large software systems based
on the experience gained in microkernel verification in the \VerisoftXT{} 
project. We discuss a number of issues that impede widespread introduction 
of formal verification in the software life-cycle process.
\end{abstract}

\section{Introduction}
In recent years, deductive program verification tools have made significant
progress. Full functional verification of individual functions written in 
real-world programming languages is practicable with reasonable effort.
Verifying large and complex software systems is also feasible, as shown in the 
L4.verified and Verisoft projects using system software as the verification 
target~\cite{Klein_EHACDEEKNSTW_10, BaumannBeckertBlasumBormer2009b}.
Naturally, verifying non-trivial software systems requires substantial
effort due to the size of the system alone. In addition, however, even
with modern specification methodologies and verification tools,
verification of a large software system suffers from scalability
issues.  Like in the software development process -- where the effort of
implementing a complete system is more than the sum of implementing
its isolated components -- the effort of specifying and verifying a
real-world system is more than the sum of verifying its components.

In this paper we report on lessons learned from the 
\VerisoftXT{} project. In the context of this project, core parts of the 
embedded hypervisor PikeOS (see \url{http://www.pikeos.com}) have been 
verified using the VCC~\cite{Cohen:TPHOLs2009-23} verification tool. 
While the size of this microkernel is several orders of magnitude smaller 
compared to, e.g., the Linux kernel, for functional verification this
can be considered to be a substantial code size. The issues presented in the paper 
are drawn from our experience with the PikeOS case study. 
Some of these issues surprised us while others were expected to occur 
from the beginning of the project on. But even for the expected problems, 
their impact on the verification effort often deviated from our anticipations. 
For example, the difficulty to verify complex algorithms and data structures 
turned out to be of little concern in the project (as the implementation of 
PikeOS avoided this complexity for good reason).

The two main conclusions from our work in the \VerisoftXT\ projects
are: firstly, verification of complex concurrent software systems can
be successfully done and, secondly, given the power of modern
verification tools, not verification but \emph{specification} is the
real bottleneck for large software systems. We argue that the latter
insight applies not only to microkernels but to most non-trivial
software system. 

The structure of the paper is as follows: In Sect.~\ref{sec:verisoftAvionics}, we
present our verification setup in the \VerisoftXT\ avionics subproject,
introducing our verification target PikeOS, the used verification tool VCC, as
well as our verification approach. Sect.~\ref{sec:lessonsLearned} presents the 
main contribution of this work, namely issues in verifying large software
systems, together with approaches to deal with them. This is followed
by related work in Sect.~\ref{sec:relatedWork} and conclusions in 
Sect.~\ref{sec:conclusion}.

\section{The \VerisoftXT\ Avionics Project}
\label{sec:verisoftAvionics}

\subsection{The PikeOS Microkernel}
\label{sec:pikeos}

PikeOS is a virtualization platform deployed
and used in industry. It consists of an L4-based microkernel acting as
paravirtualizing hypervisor and a system software component. PikeOS is written to
run on many platforms, including x86, PowerPC, MIPS, and ARM among others. While
PikeOS is able to make use of multi-processor setups, we only considered the
single-processor configuration for \VerisoftXT{} -- this decision confines 
concurrency in the kernel to preemption of user processes and interrupts. 
The PikeOS kernel is tailored to the context of embedded systems,
featuring real-time functionality and resource partitioning. The system
software component is responsible for system configuration.
Together with the system software, the PikeOS kernel provides partitioning
features that allow to virtualize several applications, such as operating
systems or run-time environments, on one CPU, where each application runs in a
secure environment with configurable access to other partitions if desired.
In order to provide real-time functionality, there are many regions within the
kernel code where execution may be preempted. Thus we have a concurrent
kernel. Moreover, the kernel is multi-threaded.

Most parts of the PikeOS kernel, especially those that are generic, are
written in~C, while other parts that are close to the hardware are necessarily
implemented in assembly. While the exact amount of assembly depends
on the architecture one works on, in our particular case, Power\-PC assembly is 
about one tenth of the codebase.

At the kernel level, the mechanisms for communication between threads are IPC,
events, and shared memory. High-level communication concepts such as
ARINC ports can be mapped onto these kernel-level mechanisms.  For a thorough
discussion of the evolution of PikeOS, see~\cite{Kaiser2007evolution}.

One consequence of using an existing and widely deployed system as verification
target is that we could not modify the code base to make it more amenable to
verification.
In addition, PikeOS has not been written with deductive verification (as used in
VCC) in mind.
On the other hand, because of the context of the application domains of PikeOS
(especially avionics), the implementation avoids overly complex (and possibly
more efficient) implementations in favor of maintainable, adaptable, and
robust code. PikeOS as a component of avionics systems has 
successfully passed DO-178B~\cite{DO-178B} evaluations, which supports this claim.

\subsection{The Verification Tool and Methodology: VCC}
\label{sec:vcc}

VCC is a deductive verification tool for concurrent C programs at the source
code level that is used to prove correctness of C implementations against
their functional specification.
The VCC toolchain allows for modular verification of concurrent C programs using
function contracts and invariants over data structures. Function contracts
are specified by pre- and postconditions. VCC is an annotation-based
system, i.e., contracts and invariants are stored as annotations within the
source code in a way that is transparent to the regular, non-verifying
compiler. 

Figure~\ref{fig:specExample} shows
the contract of a function that returns the smallest element of an array. The 
precondition ``requires'' parameter \lstinline!len! to be greater than zero and
pointer \lstinline!a! to point to an array of size \lstinline!len! in memory,
accessible to the currently executing thread (indicated by ``thread local'').
The postcondition ``ensures'' that the result is indeed the minimal element of the
array, i.e., (a)~it is less or equal than any element in the array and
(b)~actually one of the elements of the array.

\begin{figure}{\small
\begin{lstlisting}
int min(int *a, unsigned int len)
  _(requires len > 0)
  _(requires \thread_local_array(a, len))
  _(ensures \forall unsigned int i; i<len ==> \result <= a[i])
  _(ensures \exists unsigned int i; i<len && \result == a[i])
{ ... }
\end{lstlisting}}
\caption{Example VCC function contract}
\label{fig:specExample}
\end{figure}

\begin{figure}[b]
\centering
\includegraphics[width=.5\textwidth]{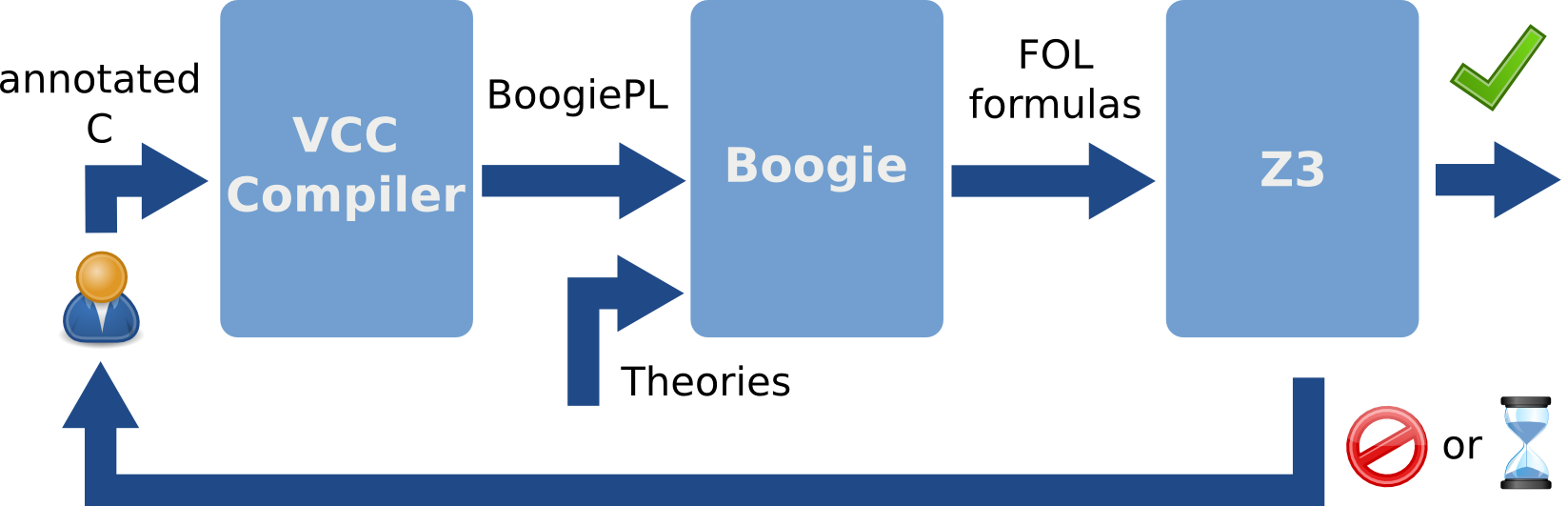}
\caption{The VCC toolchain}
\label{fig:vccToolchain}
\end{figure}

As most annotation-based verification systems today, VCC works using an
internal two-stage process. The reason for this is a better separation of
concerns and easy integration of different tools. 
As shown in Fig.~\ref{fig:vccToolchain}, the first stage of the VCC toolchain
translates the annotated C code into first-order logic via an intermediate
language called BoogiePL~\cite{DeLine-Leino05}. BoogiePL is a simple imperative
language with embedded assertions.  From this BoogiePL representation, it is
easy to generate a set of first-order logic formulas, which state that the
program satisfies the assertions. These formulas are called verification
conditions and the stage a verification condition generator~(VCG).

In the second stage, the resulting formulas are sent to an automatic theorem prover 
resp.\ SMT solver (in our case Z3~\cite{moura08z3}) together with a background 
theory capturing the semantics of C's built-in operators, etc. The prover checks 
whether the verification conditions are entailed by the background theory.  
Entailment implies that the original program is correct w.r.t.\ its specification. 
Interaction with the VCC tool is only possible (and necessary) before the
first stage of the toolchain, by providing annotations. Once sufficient
annotations have been provided (assuming the program fulfills its
specification), the proof is done automatically, hence the term
\emph{auto-active} was coined for systems following this interaction paradigm.

Other tools following the annotation-based paradigm include
Spec\#~\cite{barnett05specsharp} or Caduceus~\cite{FilliatreM04}. They are all 
based on powerful fully-automatic provers and decision procedures, and they 
support real-world programming languages such as C and~C\#.

\paragraph{Annotation-based Verification.}
Annotations can serve distinctly different purposes, though sometimes
several different ones simultaneously. Besides requirement annotations that 
assure the behavior of the program towards its environment, auto-active based
systems need auxiliary annotations to be able to prove the program correct
w.r.t.\ the requirement specification.

While one class of auxiliary annotation is needed merely for efficiency reasons 
(e.g., lemmas or intermediate assertions), another class, the \emph{essential} 
auxiliary annotations~\cite{BeckertBormerKlebanov2010} are a prerequisite for the existence of a correctness proof (e.g., loop invariants or data-structure invariants and abstractions).

Many annotations are part of the definition of module
boundaries (either implicitly or explicitly):
pre- and postconditions modularize functions, loop
invariants encapsulate functional behavior of loops, and even assertions 
within a function body may serve as modularization points (separating the part
of the function which has to establish the assertion from the part in which
the assertion is assumed to hold).
One important task in software verification is to find the right annotations at
these module boundaries, especially at the function level in case of large
software systems.

% xxx dieser absatz umgeschrieben
In theory, the strategy to come up with contracts for the functions of a large
software system is simple: first, the top-most functions in the call graph are
annotated with the weakest specification necessary to establish the requirement
specification. Then, in a top-down manner, all functions are iteratively
annotated with the weakest specification sufficient to establish correctness of
their parents. Similarly, this process could also be performed bottom-up,
starting with the strongest possible contract for the leaves of the call graph.

Unfortunately, using weakest resp.\ strongest contracts is not a good idea in 
practice. They are (a)~hard to find and (b)~hard to prove (strong contracts) resp.\ make it hard to prove the parent's correctness (weak contracts). That is, while
choosing a strong postcondition for all children of~$f$ makes verification 
of~$f$ simple, it complicates verification of the children.
So, in practice, the solution to a verification task is a set of auxiliary 
annotations that are just strong enough resp.\ just weak enough.

Moreover, the logical strength of a contract
is not its only relevant property but its syntactic form is just as important:
consider the two contracts for the function \lstinline!sqrt! that
computes (an integer approximation of) the square root of an integer shown
in Fig.~\ref{fig:alternativeSpec}. While both contracts specify the same behavior 
of \lstinline!sqrt!, one of the two contracts may be much more useful
than the other one, depending on the verification tool used and the
properties that are needed in the verification of a caller of
\lstinline!sqrt!.

\lstset{basicstyle=\ttfamily,language=C,numbers=left,xleftmargin=2em,
        numberstyle=\tiny\color{gray},mathescape=true}

\begin{figure}[tb]
\hspace*{\fill}\begin{minipage}[t]{.42\textwidth}\small
\begin{lstlisting}
int sqrt(int x)
_(requires x$\;\geq\;$0)
_(ensures 
  \result^2$\;\leq\;$x $\land$ 
  (\result+1)^2$\;>\;$x)
\end{lstlisting}
\end{minipage}\hspace*{\fill}
\begin{minipage}[t]{.57\textwidth}\small
\begin{lstlisting}
int sqrt(int x)
_(requires x$\;\geq\;$0)
_(ensures 
  $\forall$y;$\;$(y$\;\geq\;$0 $\land$ y$\;\leq\;$\result $\Rightarrow$ y^2$\;\leq\;$x) $\land$ 
  $\forall$z;$\;$(z$\;\geq\;$0 $\land$ z$\;>\;$\result $\Rightarrow$ z^2$\;>\;$x))
\end{lstlisting}
\end{minipage}\hspace*{\fill}

\caption{Two alternative contracts for the \lstinline!sqrt! function.}
\label{fig:alternativeSpec}
\end{figure}

\subsection{Verification Strategy in \VerisoftXT{}}
\label{sec:verifyingPikeOS}

For the first verification targets within PikeOS, we chose functions that were 
neither dependent on hardware features nor complex features of the programming 
or specification language (in particular, concurrency was left out). Also, 
minimal dependencies w.r.t.\ other parts of PikeOS were of advantage -- eligible
functions were identified using the call-graph of PikeOS. This  
% restriction to sequential helper functions not only
allowed for quick progress in 
verification but also reduced efforts in adapting specifications when changes in 
the specification language were made as VCC evolved. 

Next, we extended verification to sequential execution of a simple system call
that already included kernel functionality spanning all levels of the PikeOS 
microkernel, in particular PowerPC assembly~\cite{BaumannBeckertBlasumBormer2009b}. 
Here we followed the approach of Maus and Shadrin \cite{Maus08vx86, Sh12} to translate assembly 
into C code which is simulating the instruction set architecture using C data 
structures representing the hardware state.

For the verification of system calls, a first version of an abstract model of PikeOS 
according to the informal descriptions of the PikeOS kernel (e.g., the kernel's 
reference manual) was introduced. Then, using VCC's support for specification and 
verification of concurrent code, which was added in the course of \VerisoftXT, we were 
able to specify and verify preemptable system calls. This most notably included handling of 
preemption locks, interruption of threads, and 
multi-threaded concurrency. All sequentially executed parts of the system call 
between the preemption points
are represented by a transition on the abstract kernel state. Coupling 
invariants between the abstract and concrete kernel state ensure correct 
functionality of the kernel implementation. The overall correctness of PikeOS
is then expressed as a simulation theorem between the abstract model of the
kernel with the user machines and their concrete counterparts. A more detailed 
description of our verification methodology can be found in~\cite{BaumannBeckertEA2010}. 

Concerning the verification engineering work, VCC made it comparatively easy to 
argue about generic data structures like bitmap vectors and lists, as well as smaller-sized leaf 
functions in the kernel. The ubiquity of references (C pointers) to memory in the kernel 
and the memory ownership model of VCC then forced us to focus on memory 
management~\cite{BaumannBlasumBormerTverdyshev2011}. VCC verification 
also included parts of the scheduler and task tree, and numerous helper functions, in part 
including also assembly code. To give a ball-park 
figure, the overall size of the verification effort done with VCC was several thousand lines 
of code including annotations. The rest of this paper treats the microkernel as a codebase 
for real-life operating systems code, and what we encountered when using VCC on it. 
We believe the issues we met are {\em not} properties particular to {\em micro}kernels or PikeOS.

\section{Lessons Learned From Microkernel Verification}
\label{sec:lessonsLearned}

Although the list of issues presented in this section might indicate otherwise, 
our primary conclusion from the \VerisoftXT\ project is that current auto-active
verification systems are powerful enough to be successfully applied to
microkernels, i.e., complex concurrent systems.
For VCC-based verification in \VerisoftXT\, we focused on core
parts of the microkernel, and all relevant specification mechanisms could be
established. 

Independently of the size and the type of the system to be verified, the verification task
using annotation-based verification tools can be divided into the following
three phases:
\begin{enumerate}
 \item Formalization of given (informal) requirement specifications as
       program annotations.
 \item Adding auxiliary annotations to describe the boundaries and interfaces 
       of the different modules of the system
       (e.g., function contracts or loop invariants).
 \item ``Local'' verification of single modules (functions) in isolation.
\end{enumerate}
In practice, all three steps have to be performed repeatedly during several
iterations of changing annotations until a fixed-point is reached, any bugs in 
the code or the requirement specification are fixed, and verification of the
whole system succeeds.

There are several common ways to simplify the verification of large
and complex systems and make it feasible in practice: (a)~reducing the
cost of specifying and verifying a single property of a single module
(function), (b)~modularization, i.e., decomposing the verification
task by verifying one module of the system at a time, and
(c)~abstracting from details of the system's implementation and
behavior. 

All three of these concepts are addressed to a certain extent by
current deductive verification tools: (a)~Verification tools have made
a leap forward in recent years, enabling users to verify individual
functions once considered challenging with ease.  (b)~Annotation-based
verification tools like VCC make use of decomposition of the
verification task by verifying individual functions and threads in a
modular fashion. Unfortunately, in practice, the verification effort
does not scale linearly with the number of modules due to interactions
via shared data structures and common parts of the program states.
(c)~Abstraction is possible using a separate specification state and
abstract data types. But support for this is limited in
the VCC tool and methodology.

In the following, we will illustrate why verification of a system like
PikeOS still is a challenging task, despite all support by the
verification tool and methodology. Although some of the discussed
issues are more prominent for verification targets that have the
characteristics of a microkernel, they are in no way exclusive for such programs but occur to a certain extent with every large software
system. The remainder of this section is structured according to the three specification and
verification phases listed above. 

\subsection{Formalizing Requirements}

As already argued in Sect.~\ref{sec:vcc}, in theory it is not
necessary to come up with top-level contracts adapted to the (often
informal) requirement specification of the system, as the strongest
possible contract must always suffice. However, in practice, the
strongest contract not only makes verification more complicated, but
it also obscures the intention behind the behavior of the system. In
addition, regardless of the strength of a contract, the user has to
find the right kind of abstraction when formalizing informal
requirements. For these reasons, formal top-level contracts must take the
informal requirements and other system documentation into
consideration.

\paragraph{Issue: Implicit Behavior in Informal Specifications.}
\label{sec:codeToSpec}

There exists rich official material in form of end-user documentation and
requirement engineering documentation for PikeOS. These are however focused on the strict separation of platforms, architecture, and kernel (which is justified from a maintenance perspective) and is thus of limited use for finding annotations for functional verification.
\subpara{Example}
An example for user-level documentation keeping concurrency effects implicit is the informal specification of a system call that changes
the priority of a thread (confined to the user-defined value MCP, the ``maximum
controlled priority'' of a task), taken from the kernel reference manual:
{\it
``This function sets the current thread's priority to newprio.
Invalid or too high priorities are limited to the caller's task MCP.
Upon success, a call to this function returns the current thread's priority
before setting it to newprio.''
}
However, this system call is preemptable, and if during a preemption another
user thread has changed the thread's priority before the function's return
value is assigned, then the ``old priority'' returned might not be what a naive observer might expect who neglects that the system is concurrent~\cite{BaumannBeckertBlasumBormer2009b}.%
\subpara{Approaches to Resolving the Issue}  
Ideally, to simplify the transition from the more informal existing system 
documentation, a first step would be a specification mechanism that allows to 
write down the intention of the programmer respectively system architect 
explicitly, although without the need of a complex formal specification.
Preferably, these documents have to be of value in the software development
process besides formal verification. 
An example of such precise specification of the expected system behavior are test cases.

In this regard, also documents supporting certification measures are helpful: 
the DO-178 avionics certification requires, e.g., descriptions of 
concurrency and reentrance analyses. Also the Common Criteria for Information 
Technology Security give suggestions to system architects in how to structure an 
architecture of a system and how to describe the security properties of a system in 
terms of domain separation, self-protection and non-bypassibility.

Besides existing documentation, code inspection is a practicable way to get a 
first version of the auxiliary annotations needed for the verification
of the requirement specification. However, this involves the risk of repeating
mistakes in the specification that were already introduced in the
implementation.

\paragraph{Issue: No Syntactic Distinction Between Different Kinds of
Annotations.} 

Besides the requirement specification, there are two kinds of auxiliary
annotations needed for verification with VCC, as already stated in
Sect.~\ref{sec:vcc}.
However, in the case of VCC, there is no clearly visible distinction between
auxiliary and requirement specification and between essential and non-essential annotations 
added for performance reasons. 
We argue in~\cite{BeckertBormerKlebanov2010} that it is extremely important for
the user to have knowledge about which kind of annotations are essential for the
verification system as without that knowledge they may continue to add the wrong
annotations in case of a failed proof attempt.

Moreover, we claim that requirement and auxiliary annotations must be
syntactically distinguished. That makes specifications clearer and easier to
read and understand. In certification processes it is indispensable to
have a very clear understanding of which annotations form the
requirement specification that has been verified.
Also, managing annotations in case of software evolution is more difficult when
no clear distinction is given: while auxiliary annotations may be changed
or removed at will, requirements have to stay unchanged. As a consequence,
without this distinction, maintainability of the annotations suffers, e.g., 
by having to separate requirements from auxiliary annotations in such a case.
\subpara{Example}
When VCC verifies an inline function, a full specification of pre- and postconditions is needed. However, the specification of the inline function is an implementation detail and not part of the external requirements. 
Inline functions are common for our verification target for efficiency reasons,
one example is the function to restore interrupts described
in~\cite{BaumannBeckertBlasumBormer2009b}.
\subpara{Approaches to Resolving the Issue}
It is preferable to keep requirement and auxiliary annotations separate, e.g.,
requirement annotations in the header file and auxiliary
annotations in the C~source file. Where no such separation is possible, keywords
(in the style of visibility modifiers \texttt{public} and
\texttt{private}) should be used.

\subsection{Adding Auxiliary Annotations}
\label{sec:addingAnn}

After the requirements have been formalized as annotations,
a multitude of essential auxiliary annotations have to be added before
even the first verification attempt can begin. Among these auxiliary
annotations are function contracts, loop invariants, as well as data structure
invariants. In most cases, this initial set of auxiliary annotations is not
sufficient to prove the part of the system considered correct and the
annotations have to be adapted in several iterations, for reasons
further explained below.

\subsubsection{Modularization}

Modularization helps to reduce verification effort by decomposing the 
verification task. For the modular verification of sequential programs with VCC,
a function is verified using the contracts of all the  functions it calls -- 
instead of their 
implementations. This design choice relieves the load on the automated prover at
the expense of having to write (auxiliary) function contracts for all functions.

For this reason, using current auto-active verification tools, the bottleneck
of verification is
how to find the right auxiliary specifications and specifications of interfaces 
between modules. In addition, architectures such as microkernels feature some inherent 
characteristics that restrict the extent to which the specification task can be 
modularized.

In the case of the PikeOS microkernel, implementations of single C functions
are deliberately kept simple to facilitate maintainability and certification
measures -- the functionality of the whole kernel is rather implemented by
interaction of many of these small functions, operating on common data structures.
Microkernels, and more generally, all operating systems, have to keep track of 
the overall system's state, resulting in relatively large and complex data 
structures on which many of the kernels functions operate conjointly. This 
amount of interdependencies has an impact on the following issues.

\paragraph{Issue: Entangled Specifications.}
Function specifications have strong dependencies on each other. 
Finding the right annotations for a 
single function requires the verification engineer
to consider several functions at once, due to these dependencies. Good feedback
of the verification tool in case of a failed verification attempt is essential.
Feedback given by annotation-based verification tools so far only
focuses on the function currently being verified. This allows the user to
pinpoint 
and fix bugs in the specification or program respectively to change auxiliary 
annotations, e.g., loop invariants or contracts for called functions. For the 
verification of single functions, this tool support is sufficient. However,
current specification methodologies are lacking w.r.t.\ providing help to the user in 
case of analyzing problems with interdependent specifications.
\subpara{Example}
Already the relatively simple examples given in~\cite{BaumannBlasumBormerTverdyshev2011,BaumannBeckertBlasumBormer2009b} both have a call graph of depth three, showing the interdependency of the annotations between callers and callees.
\subpara{Approaches to Resolving the Issue}
One solution to this issue, we considered in~\cite{BeckertBormerMerzSinz2011},
is to provide early feedback when starting to specify a software system. Our
approach is based on the combination of deductive verification as in VCC with
software bounded model checking (SBMC) using the LLBMC
tool~\cite{SFM2010aprecisememorymodel}.
For this, annotations written in VCC's specification language are translated 
into assertions that can be checked by LLBMC (i.e., boolean C~expressions 
extended with some features specific to LLBMC).
The SBMC procedure then allows to check these assertions without the need to
provide any additional auxiliary annotations as in case with VCC, e.g.,
functions are
inlined and loops unrolled instead of modularized using annotations.
However, in contrast to deductive verification, the  
number of loop iterations resp.\ the function invocation depth is bounded.
If no assertion is violated within the given bounds, this does not imply that
the program adheres to its specification, as a bug may still occur, e.g., in a
loop iteration outside the given bound.
That is, LLBMC does not provide a full proof (that is left to VCC) but a quick
check that can point the user to problems in the annotations early on and thus
avoid unnecessary VCC proof attempts. 

\begin{figure}[tb]%
\centering
\subfloat[][]{\label{fig:annotationLoop}\includegraphics[scale=0.45]{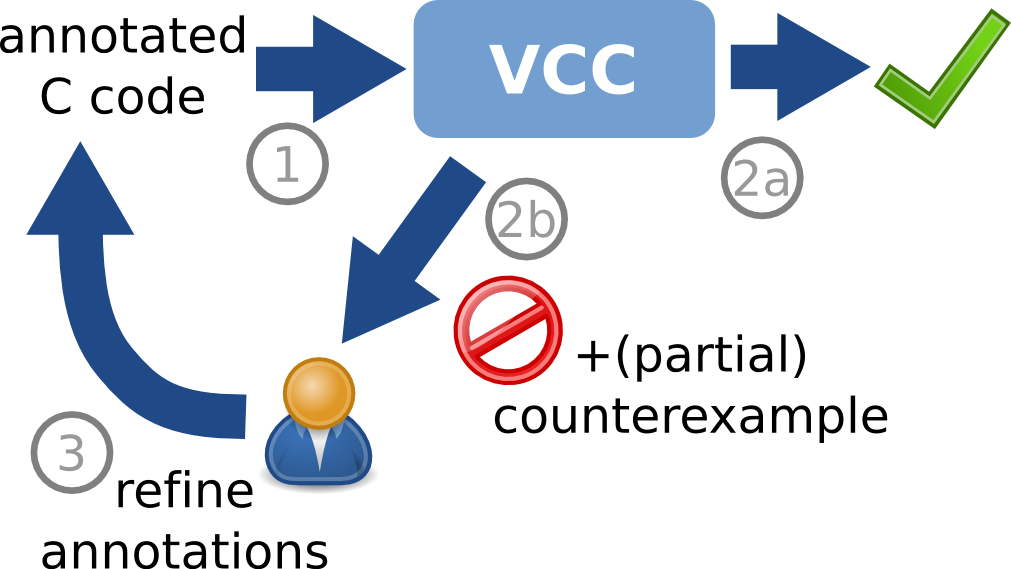}
} %
\quad%
\vline%
\quad%
\subfloat[][]{\label{fig:cegmarLoop}\includegraphics[scale=0.45]{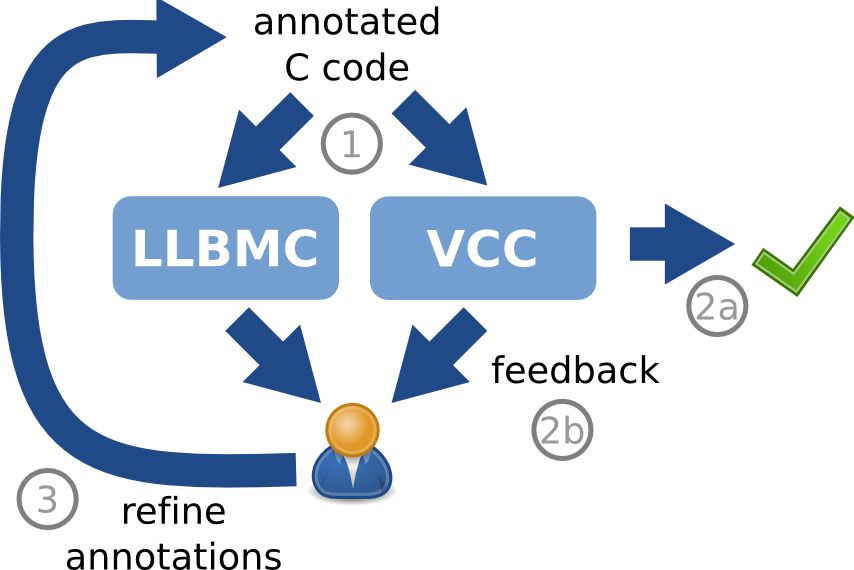}
} %
\caption{\protect\subref{fig:annotationLoop} Normal VCC workflow 
and~\protect\subref{fig:cegmarLoop} counterexample guided 
\emph{manual annotation} refinement (CEGMAR).}%
\label{fig:cont}%
\end{figure}

A comparison of the regular VCC workflow with the SBMC-supported variant is
shown in Fig.~\ref{fig:cont}. Using VCC alone (Fig.~\ref{fig:annotationLoop}),
starting from the sufficiently annotated C code, VCC is invoked (step~1). If VCC
verifies the program to fulfill its specification (step~2a), the process ends.
More often, though, verification fails (step~2b) and the user has to change 
annotations using the counterexample provided by VCC (step~3).

Figure~\ref{fig:cegmarLoop} shows the same process, this time with guidance by
LLBMC: the annotated C code is given to both VCC and LLBMC in step 1.
In case LLBMC reports a violated assertion (step 2b), the user
is given a concrete trace through the program that leads to this violation,
which is very helpful in finding either bugs in the implementation or
specification.

\paragraph{Issue: Module Granularity.}
Dependencies between functions obfuscate module boundaries and thus
make finding the right module interfaces difficult. 
In addition, even if larger modules can be identified, there is no particular 
support for a hierarchical modularization in annotation-based verification 
systems and for specifying properties of such a larger module, e.g., on the system 
architecture level. 

Also, by the VCC methodology, module boundaries are fixed by having to provide
function contracts for all functions -- this implies that every function is
also a module. However, some function boundaries of the software system might
also be chosen for reasons other than clear separation of disjoint functionality
of different modules. Ideally, for those functions
of a large software system that are not part of the externally visible part of
the system's interface, such function contracts could be omitted -- especially
for small library functions that occur at the leaves of the system's call graph.
\subpara{Example}
VCC simply treats each function as a module. Conversely, modules in PikeOS are 
not restricted to one function.
\subpara{Approaches to Resolving the Issue}
Inlining of function calls would help to make function contracts obsolete for 
those functions of the system that are not a module on its own but only part of
some larger functionality. This would already moderate the issue of imposing
fixed module boundaries for functions in some cases. 

%On the sub-function level, VCC allows to define blocks of code with separate
%contracts inside a function body. On a larger scale a similar mechanism would be
%helpful, e.g., to define and give specifications for function hierarchies.

\subsubsection{Abstraction}

Besides modularization, abstraction is another important instrument to handle
verification of large software systems. Good abstraction of the functional
behavior of a system helps to focus on important details of the functionality,
and allows for clear and succinct specifications. Poorly chosen abstractions
may complicate verification up to making it impossible to verify the system at
all -- which abstraction is appropriate not only depends on the system
properties to be specified but also on how well the verification tool used is
able to reason about it.

\paragraph{Issue: Finding the Right Abstraction.}

To find the right abstraction for data structures, analyzing the source code 
alone is often not sufficient in practice. Complexity of the 
implementation of a single data structure is secondary when considering 
verification effort. Gathering the important properties of the data structure is 
crucial in order to find the right abstraction. Which of these properties are 
needed and relied upon in the software depends on its usage in the functions of
the system. While some techniques such as CEGAR exist that may help in some
cases in finding the right abstractions, these functions are not sufficiently
supported in current deductive annotation-based systems.

Again, multiple dependencies between the functions that all operate on (parts
of) the same data structure make it hard to find the right abstraction.
Besides better support from verification tools, in order to come up with the 
right abstractions, information from system developers and architects is vital.
\subpara{Example}
The concept of representing chunks of memory as {\em objects} is one abstraction 
provided by VCC. On top of this notion of objects, the concept of 
{\em ownership sets} (consisting of objects), as well as the concept to define 
specification {\em maps} (with objects as their domain) is provided by VCC as two 
other abstraction mechanisms. Both of these abstraction features are used 
in~\cite{BaumannBlasumBormerTverdyshev2011} where ownership sets provide a 
grouping of objects of the same type to ``object managers'', whereas a map expresses 
separation of objects. 

\paragraph{Issue: No Language Support for Abstractions.}

Abstraction in the VCC tool is mainly achieved by using specification
(``ghost'') state together with built-in abstract data types like maps or sets,
where the abstract state is related to the concrete program state with the help
of coupling invariants. 

One advantage of having a similar notation for the ghost data structures and
statements compared to real C is that programmers are already familiar with it.
However, at times, such notations are still too close to the source code and
better suited alternatives exist -- in general, mechanisms to specify
abstractions have to be more flexible in order to be able to choose the right
abstraction methodology for the task at hand. 
\subpara{Examples}
Lists (used in~\cite{BaumannBlasumBormerTverdyshev2011}) would benefit from a recursive specification. Another example where a recursive specification would have been more natural is the PikeOS task tree~\cite{Kaiser2007evolution} and functions using it.
\subpara{Approaches to Resolving the Issue}
Concerning data abstraction, user-defined abstract data types were
already introduced into the VCC methodology, however, there is a large amount
of established formalisms, like CASL, that should be taken into further
consideration when extending the specification language.

For control abstraction, many established formalisms exist that could be
used for one of the abstraction layers on top of the code, e.g., CSP or
abstract state machines.
Also, a built-in refinement mechanism is needed to connect the different
abstraction levels.

\subsection{Local Verification}

Deductive software verification tools have improved in recent years to a degree
that full functional verification of individual functions written in 
real-world programming languages is practicable with reasonable effort.
This is demonstrated, e.g., by the results of recent verification 
competitions (e.g.,~\cite{COSTCompetition2011}), where selected software
verification problems are solved with limited time resources by teams ranging
from verification tool developers to regular tool users.

In the following, we consider for small scale verification the verification of
a single function against a \emph{given} informal requirement specification.
This includes formalizing the requirement specification in a way that is
suitable to the verification methodology at hand, as well as coming up with all
auxiliary specifications necessary to verify the function correct w.r.t.\ its
contract.

\paragraph{Issue: Support in Finding Auxiliary Annotations.}

Finding the right auxiliary annotations for local verification is a complicated
task. One issue, finding the right essential annotations at module boundaries,
has already been mentioned in Sect.~\ref{sec:addingAnn} -- besides the obvious
function contracts, also loop invariants belong to this group. Even if, after
some iterations, these essential annotations are appropriate for verifying the
function in question, finding auxiliary annotations between
modularization points remains to be done.

As functional program verification is undecidable, it is not
surprising that coming up with the right annotations is non-trivial. Until now,
there is little support by the tools in finding the right auxiliary
annotations: in case of performance problems of the prover, the user can
inspect the proof process and investigate which annotation takes up
most of the time. In addition, statistics of the underlying SMT solver are
reported, e.g., the number of quantifier instantiations so far, in order to find
bottlenecks. In case of too weak essential annotations, VCC produces a
counterexample that helps to find the missing or wrong annotations. 

If none of these facilities provide the right clue, the user is left with
``debugging'' the verification state by inserting additional assertions to
further split the proof and isolate those parts of annotations that are
difficult for the automatic prover.
\subpara{Example}
At some points the behavior of the verification system is unexpected and
experience in using the verification system is crucial: for example, the
assertion \lstinline!\exists int i; i == 1! cannot be verified by VCC without
providing another essential annotation in form of a \emph{trigger}. Another
stumbling block is that not only is the ordering of assertions of importance
in finding a proof, but also is the ordering of loop invariant annotations.
Also, for performance reasons, some information about the proof state is
discarded when calling other functions and the user has to explicitly state
which information to keep after the function call for the verification of the
remaining function.

\paragraph{Issue: Amount of Annotations.} Although small scale verification is
possible using current verification tools and
methodologies with reasonable effort, there is still room for improvements.
Most notably, the large amount and high verbosity of annotations is one issue
when using auto-active verification tools. In case of VCC, the specification 
language has been optimized several times to reduce annotation overhead.%
\subpara{Example}
For the functional verification of parts of the memory 
manager in PikeOS~\cite{BaumannBlasumBormerTverdyshev2011} the overhead of 
specification compared to the code compared by lines was still 5:1, using VCC as
of Feb.~11,~2011.
\subpara{Approaches to Resolving the Issue}
An obvious way to reduce the amount of annotations necessary, besides improving
the performance of the SMT solver, is to identify common specification patterns
and create new specification constructs as abbreviation. Another possibility is 
to choose defaults for specification constructs that cover the most frequent
explicit specifications.

Note that in case of C, function declarations are given in separate header files
and VCC takes advantage of this by allowing to annotate these declarations with
the function contract. This separates the interface specification 
from the auxiliary specification in the function body but
at the same time impairs usability of the verification tool: either the user 
does not have the contract and the implementation visible side by side when
verifying or inspecting the function, or he/she duplicates resp.\ moves the
contract to the implementation, incurring additional overhead to synchronize
the header file with the implementation.

But even with the interface specification taken care of, the problem of a large amount of auxiliary annotations in the
function body remains. Again, support of a special verification development 
environment could help the user to keep track of both implementation and 
specification. Parts of the annotations irrelevant for the user in 
understanding the specification could be hidden, given a heuristic that 
determines the relevancy of an annotation. Information that could be taken into
account for such a heuristic is a similarity measure on annotations resp.\
formulas (cf.~\cite{GrebingDA2012}) -- this would allow to hide all but one
annotation in a group of similar annotations.

Also, adapting the technique of code refactoring to specifications could be used
to reduce the amount and improve readability and maintainability of annotations,
e.g., by factoring out often used blocks of specifications in a specification 
function.

One option to reduce amount of required auxiliary annotations
is to shift user interaction towards the proof construction stage, as
done in interactive provers. Annotations would only contain the main properties
and insights about the implementation (e.g., loop invariants). Proof 
guidance (e.g., quantifier instantiations) would be done interactively (the 
information provided by the user can be stored in an explicit proof 
object for reuse, as in, e.g., the \KeY{} verification tool~\cite{KeYBook2007}).

\subsection{Handling Software Evolution}

While the one-time effort in verifying a large software system is already high,
this does not yet include expenditures for re-verification in case of software
evolution. In order to integrate functional software verification into the
software development process as part of the software quality control, costs for
re-verification have to be reduced, e.g., by re-using as many annotations as
possible.

\paragraph{Issue: Change Management.}
Another issue that arises when verifying an implementation in productive use is 
that it is constantly evolving, for example, to satisfy changing requirements 
from users of the software. 
To cope with these changes, when deciding on the frequency of applying code 
updates, one has to find a balance between costs for re-verification and 
the benefits of improvements in the system that might simplify specification
and verification.

One possibility is to constantly apply the changes that occur in the production 
code to the source snapshot that is used as the verification target. While some
proofs are not affected by such small changes, it is in general still necessary 
to adapt code annotations and verification proofs.
The other possibility is to fix a version of the source code for verification. 
Then, to get verified properties for the code used in the actual product, the 
annotations of the verification target have to be adapted in one big leap to
the current implementation at the end of the project. 
\subpara{Example}
In \VerisoftXT{}, in the beginning, we constantly applied changes occurring in the production code (at the expense of time available for the actual verification progress). At a later point, no more updates were done the verification target.
To support change management, a test harness was used for regression testing to
check whether previously completed proofs were still valid or had to be redone.

\section{Related Work}
\label{sec:relatedWork}

Two recent large scale verification projects relevant for our work 
are the L4.verified project~\cite{Klein_EHACDEEKNSTW_10} and the Hyper-V 
sub-project~\cite{springerlink:10.1007/978-3-642-05089-3_51} within 
\VerisoftXT{}. Compared to PikeOS, the Hyper-V source code is roughly
one order of magnitude larger with about 100kLOC. Also, Hyper-V makes use of a
multi-core hardware setup. Similar to the setup of PikeOS, 
the code base of the hypervisor was not adapted to simplify verification. 
Concerning the amount of annotations per line of source code needed, a maximum
ratio of three to one is reported~\cite{Tobies2011}.

In another sub-project of \VerisoftXT{}, the
functional verification of a simple version of a hypervisor has been accomplished
showing feasibility of auto-active verification for system software and which
has been used to improve the VCC tool for further use in the \VerisoftXT{} 
project. Specification effort mentioned in~\cite{AlkassarEtAl2010} matches the
ratio reported for Hyper-V: the implementation consisted about 2.5k C code
tokens compared to the 7.7k annotation tokens needed.
Based on experience gathered within verification of those hypervisors, the VCC
methodology and tool chain has been improved with the goal to increase 
performance of the automated prover and especially to guarantee fast response
times in case of failed verification attempts, as this is the common case in
software verification.

In the L4.verified project, an approach was taken that differs in various ways from the one used 
for the PikeOS and Hyper-V verification in \VerisoftXT.
Besides using the interactive theorem 
prover Isabelle/HOL for specification and verification of the seL4 microkernel,
the implementation of the kernel was written from scratch for the project and
adapted to suit verification, if needed. Also, an abstraction layer in form of a
Haskell implementation of the kernel has been reported to be of great help
(instead of directly verifying the equivalent C implementation). Both 
implementations in Haskell and C together amount to approx.~15kLOC, while the 
Isabelle script used in verification spans 200kLOC~\cite{Klein_EHACDEEKNSTW_10}.

\paragraph{Improving the Verification Process.}
To simplify the specification and verification process, being able to specify
abstractions of code-level properties is important. There are many formalisms
to provide control and data abstraction -- of those, we already mentioned the
Common Algebraic Specification Language (CASL), as well as Communicating
Sequential Processes (CSP) or Abstract State Machines (ASM). Other approaches to
improve usability of the specification languages take their cues from software
engineering: for example specification
patterns~\cite{Bianculli:2012:SPR:2337223.2337341} or specification
refactoring~\cite{Goldstein2007, Hull2010} are useful instruments
to facilitate writing and maintaining specifications.

A combination of interactive verification and auto-active tools has been used
for the verification of checkers for complex algorithmic implementations
in~\cite{DBLP:conf/cav/AlkassarBMR11}, taking advantage of VCC for code level
verification and Isabelle/HOL to verify mathematical properties.
Also, the Isabelle/HOL framework has been used in the HOL-Boogie
tool~\cite{Bohme:2008:HIP:1459784.1459802}, which has been developed to
interactively perform complex proofs that VCC cannot handle.

\section{Conclusion}
\label{sec:conclusion}

Deductive program verification tools have made significant progress in recent
years that allows to apply them to complex concurrent software. 
However, to be able to efficiently
verify large software, further improvements are necessary. As the modules
that can be verified using current tools get larger and more complicated, it
becomes apparent that the process of finding the right specification is now
the bottleneck.

In this paper, we have described issues in software verification that occurred
during the verification of PikeOS within the scope of the \VerisoftXT{} project and that contribute to the problem of coming up with sufficient code annotations. For most issues, we presented ideas on how to resolve the shortcomings.
In general, in order to handle verification of large software systems
efficiently, we claim that better support from the verification tool is needed 
to get from verification
of individual functions to verification of whole software systems. This would have 
to include support in finding the right modularization and abstraction.
A first step in this direction is to provide the user with feedback 
in case of interdependent function specifications, so that mismatching
contracts 
are discovered early in the specification
process~\cite{BeckertBormerMerzSinz2011}. 

Besides annotation-based specifications, which are well suited for describing
implementations at source-code level, there is a need for further specification
constructs tailored to describing the system properties at higher levels 
of abstraction. Amongst others, these formalisms should allow to integrate 
knowledge of system developers even before starting the specification process 
at code level. 
At the code level, we propose to make use of established 
formalisms for abstract data types like CASL, to be able to specify common 
implementation data structures in a compact 
manner~\cite{BeckertBormerKlebanov2010}.

Compared to verification of the academic system in the first phase of 
Verisoft, where Isabelle/HOL was applied, in the successor project \VerisoftXT{}, 
verification was performed using the auto-active methodology of the VCC tool. 
VCC was developed within \VerisoftXT{} with the goal of providing a high degree
of automation in verifying functional properties of a program, which allows
to scale verification to large software systems like the Hyper-V or PikeOS.

Some of the challenges presented in this paper are to a certain degree a 
consequence of this auto-active verification approach and the design decisions 
made in VCC, like the inflexible module granularity or the amount of annotations
needed. Other issues mentioned affect both auto-active tools and
interactive verification systems: entangled module specifications are of concern 
to all modular verification methodologies, also change management is a major 
issue in verifying evolving, real-world software.

For some of these issues shared by both verification approaches alike, 
interactive proof construction offers significant advantages -- amongst others,
guiding the prover in finding a proof (similar to finding sufficient annotations
in case of VCC) is supported by being able to inspect the current proof state.
Interactive proof systems like Isabelle/HOL also are more flexible 
concerning specification formalisms.

When a successful formal verification of complex and large software systems at code level is desired, we believe
a combination of both interactive and auto-active specification and verification 
approaches is promising -- similar to verification of certifying algorithms 
using VCC, together with Isabelle/HOL as shown in~\cite{DBLP:conf/cav/AlkassarBMR11}.
Using this combination, auto-active tools would allow for efficient verification
of source-code level properties closer to the hardware, the resulting functional
properties of the code abstracting from implementation details -- whereas 
interactive verification systems may be used to verify complex system properties
on a suitable, user-defined abstraction of the system.

\paragraph{Acknowledgment}

We thank the anonymous reviewers for helpful comments.
We also thank our previous \VerisoftXT{} co-workers (in particular those who worked with VCC on PikeOS, i.e., J\'{e}r\^{o}me Creci, Dilyana Dimova and Markus Wagner).

\bibliographystyle{eptcs}
\bibliography{main}

\begin{thebibliography}{10}
\providecommand{\bibitemdeclare}[2]{}
\providecommand{\surnamestart}{}
\providecommand{\surnameend}{}
\providecommand{\urlprefix}{Available at }
\providecommand{\url}[1]{\texttt{#1}}
\providecommand{\href}[2]{\texttt{#2}}
\providecommand{\urlalt}[2]{\href{#1}{#2}}
\providecommand{\doi}[1]{doi:\urlalt{http://dx.doi.org/#1}{#1}}
\providecommand{\bibinfo}[2]{#2}

\bibitemdeclare{inproceedings}{DBLP:conf/cav/AlkassarBMR11}
\bibitem{DBLP:conf/cav/AlkassarBMR11}
\bibinfo{author}{Eyad \surnamestart Alkassar\surnameend},
  \bibinfo{author}{Sascha \surnamestart B{\"o}hme\surnameend},
  \bibinfo{author}{Kurt \surnamestart Mehlhorn\surnameend} \&
  \bibinfo{author}{Christine \surnamestart Rizkallah\surnameend}
  (\bibinfo{year}{2011}): \emph{\bibinfo{title}{Verification of Certifying
  Computations}}.
\newblock In \bibinfo{editor}{Ganesh \surnamestart Gopalakrishnan\surnameend}
  \& \bibinfo{editor}{Shaz \surnamestart Qadeer\surnameend}, editors: {\sl
  \bibinfo{booktitle}{CAV}}, \bibinfo{series}{LNCS 6806},
  \bibinfo{publisher}{Springer}, pp. \bibinfo{pages}{67--82},
  \doi{10.1007/978-3-642-22110-1\_7}.

\bibitemdeclare{incollection}{AlkassarEtAl2010}
\bibitem{AlkassarEtAl2010}
\bibinfo{author}{Eyad \surnamestart Alkassar\surnameend}, \bibinfo{author}{Mark
  \surnamestart Hillebrand\surnameend}, \bibinfo{author}{Wolfgang \surnamestart
  Paul\surnameend} \& \bibinfo{author}{Elena \surnamestart Petrova\surnameend}
  (\bibinfo{year}{2010}): \emph{\bibinfo{title}{Automated Verification of a
  Small Hypervisor}}.
\newblock In: {\sl \bibinfo{booktitle}{Verified Software: Theories, Tools,
  Experiments}}, {\sl \bibinfo{series}{LNCS}} \bibinfo{volume}{6217},
  \bibinfo{publisher}{Springer}, pp. \bibinfo{pages}{40--54},
  \doi{10.1007/978-3-642-15057-9\_3}.

\bibitemdeclare{inproceedings}{barnett05specsharp}
\bibitem{barnett05specsharp}
\bibinfo{author}{Mike \surnamestart Barnett\surnameend},
  \bibinfo{author}{K.~Rustan~M. \surnamestart Leino\surnameend} \&
  \bibinfo{author}{Wolfram \surnamestart Schulte\surnameend}
  (\bibinfo{year}{2005}): \emph{\bibinfo{title}{The {Spec\#} Programming
  System: An Overview}}.
\newblock In: {\sl \bibinfo{booktitle}{Construction and Analysis of Safe,
  Secure, and Interoperable Smart Devices (CASSIS), International Workshop,
  2004, Marseille, France, Revised Selected Papers}}, \bibinfo{series}{LNCS
  3362}, \bibinfo{publisher}{Springer}, pp. \bibinfo{pages}{49--69},
  \doi{10.1007/978-3-540-30569-9\_3}.

\bibitemdeclare{inproceedings}{BaumannBlasumBormerTverdyshev2011}
\bibitem{BaumannBlasumBormerTverdyshev2011}
\bibinfo{author}{C.~\surnamestart Baumann\surnameend},
  \bibinfo{author}{H.~\surnamestart Blasum\surnameend},
  \bibinfo{author}{T.~\surnamestart Bormer\surnameend} \&
  \bibinfo{author}{S.~\surnamestart Tverdyshev\surnameend}
  (\bibinfo{year}{2011}): \emph{\bibinfo{title}{{Proving Memory Separation in a
  Microkernel by Code Level Verification}}}.
\newblock In \bibinfo{editor}{Wilfried \surnamestart Steiner\surnameend} \&
  \bibinfo{editor}{Roman \surnamestart Obermaisser\surnameend}, editors: {\sl
  \bibinfo{booktitle}{{1st International Workshop on Architectures and
  Applications for Mixed-Criticality Systems (AMICS 2011)}}},
  \bibinfo{publisher}{IEEE Computer Society}, \bibinfo{address}{Newport Beach,
  CA, USA}, \doi{10.1109/ISORCW.2011.14}.

\bibitemdeclare{inproceedings}{BaumannBeckertBlasumBormer2009b}
\bibitem{BaumannBeckertBlasumBormer2009b}
\bibinfo{author}{Christoph \surnamestart Baumann\surnameend},
  \bibinfo{author}{Bernhard \surnamestart Beckert\surnameend},
  \bibinfo{author}{Holger \surnamestart Blasum\surnameend} \&
  \bibinfo{author}{Thorsten \surnamestart Bormer\surnameend}
  (\bibinfo{year}{2009}): \emph{\bibinfo{title}{{Formal Verification of a
  Microkernel Used in Dependable Software Systems}}}.
\newblock In \bibinfo{editor}{Bettina \surnamestart Buth\surnameend},
  \bibinfo{editor}{Gerd \surnamestart Rabe\surnameend} \& \bibinfo{editor}{Till
  \surnamestart Seyfarth\surnameend}, editors: {\sl
  \bibinfo{booktitle}{{SAFECOMP'09}}}, \bibinfo{series}{{LNCS 5775}},
  \bibinfo{publisher}{Springer}, pp. \bibinfo{pages}{187--200},
  \doi{10.1007/978-3-642-04468-7\_16}.

\bibitemdeclare{inproceedings}{BaumannBeckertEA2010}
\bibitem{BaumannBeckertEA2010}
\bibinfo{author}{Christoph \surnamestart Baumann\surnameend},
  \bibinfo{author}{Bernhard \surnamestart Beckert\surnameend},
  \bibinfo{author}{Holger \surnamestart Blasum\surnameend} \&
  \bibinfo{author}{Thorsten \surnamestart Bormer\surnameend}
  (\bibinfo{year}{2010}): \emph{\bibinfo{title}{{Ingredients of Operating
  System Correctness}}}.
\newblock In: {\sl \bibinfo{booktitle}{{Proceedings, embedded world 2010
  Conference, Nuremberg, Germany}}}.
\newblock \bibinfo{note}{Available at
  \url{http://formal.iti.kit.edu/beckert/pub/embeddedworld2010.pdf}}.

\bibitemdeclare{incollection}{BeckertBormerKlebanov2010}
\bibitem{BeckertBormerKlebanov2010}
\bibinfo{author}{Bernhard \surnamestart Beckert\surnameend},
  \bibinfo{author}{Thorsten \surnamestart Bormer\surnameend} \&
  \bibinfo{author}{Vladimir \surnamestart Klebanov\surnameend}
  (\bibinfo{year}{2012}): \emph{\bibinfo{title}{Improving the Usability of
  Specification Languages and Methods for Annotation-Based Verification}}.
\newblock In \bibinfo{editor}{Bernhard \surnamestart Aichernig\surnameend},
  \bibinfo{editor}{Frank \surnamestart de~Boer\surnameend} \&
  \bibinfo{editor}{Marcello \surnamestart Bonsangue\surnameend}, editors: {\sl
  \bibinfo{booktitle}{Formal Methods for Components and Objects}}, {\sl
  \bibinfo{series}{LNCS}} \bibinfo{volume}{6957}, pp. \bibinfo{pages}{61--79},
  \doi{10.1007/978-3-642-25271-6\_4}.

\bibitemdeclare{inproceedings}{BeckertBormerMerzSinz2011}
\bibitem{BeckertBormerMerzSinz2011}
\bibinfo{author}{Bernhard \surnamestart Beckert\surnameend},
  \bibinfo{author}{Thorsten \surnamestart Bormer\surnameend},
  \bibinfo{author}{Florian \surnamestart Merz\surnameend} \&
  \bibinfo{author}{Carsten \surnamestart Sinz\surnameend}
  (\bibinfo{year}{2011}): \emph{\bibinfo{title}{Integration of Bounded Model
  Checking and Deductive Verification}}.
\newblock In: {\sl \bibinfo{booktitle}{FoVeOOS'11}}, pp.
  \bibinfo{pages}{86--104}, \doi{10.1007/978-3-642-31762-0\_7}.

\bibitemdeclare{book}{KeYBook2007}
\bibitem{KeYBook2007}
\bibinfo{editor}{Bernhard \surnamestart Beckert\surnameend},
  \bibinfo{editor}{Reiner \surnamestart H\"ahnle\surnameend} \&
  \bibinfo{editor}{Peter~H. \surnamestart Schmitt\surnameend}, editors
  (\bibinfo{year}{2007}): \emph{\bibinfo{title}{Verification of Object-Oriented
  Software: The {KeY} Approach}}.
\newblock \bibinfo{series}{LNCS 4334}, \bibinfo{publisher}{Springer-Verlag},
  \doi{10.1007/978-3-540-69061-0}.

\bibitemdeclare{inproceedings}{Bianculli:2012:SPR:2337223.2337341}
\bibitem{Bianculli:2012:SPR:2337223.2337341}
\bibinfo{author}{Domenico \surnamestart Bianculli\surnameend},
  \bibinfo{author}{Carlo \surnamestart Ghezzi\surnameend},
  \bibinfo{author}{Cesare \surnamestart Pautasso\surnameend} \&
  \bibinfo{author}{Patrick \surnamestart Senti\surnameend}
  (\bibinfo{year}{2012}): \emph{\bibinfo{title}{Specification patterns from
  research to industry: a case study in service-based applications}}.
\newblock In: {\sl \bibinfo{booktitle}{Proceedings of the 2012 International
  Conference on Software Engineering}}, \bibinfo{series}{ICSE 2012},
  \bibinfo{publisher}{IEEE Press}, \bibinfo{address}{Piscataway, NJ, USA}, pp.
  \bibinfo{pages}{968--976}, \doi{10.1109/ICSE.2012.6227125}.

\bibitemdeclare{inproceedings}{Bohme:2008:HIP:1459784.1459802}
\bibitem{Bohme:2008:HIP:1459784.1459802}
\bibinfo{author}{Sascha \surnamestart B\"{o}hme\surnameend},
  \bibinfo{author}{K.~Rustan \surnamestart Leino\surnameend} \&
  \bibinfo{author}{Burkhart \surnamestart Wolff\surnameend}
  (\bibinfo{year}{2008}): \emph{\bibinfo{title}{HOL-Boogie -- An Interactive
  Prover for the Boogie Program-Verifier}}.
\newblock In: {\sl \bibinfo{booktitle}{Proc., 21st Int. Conf. on Theorem
  Proving in Higher Order Logics}}, \bibinfo{publisher}{Springer}, pp.
  \bibinfo{pages}{150--166}, \doi{10.1007/978-3-540-71067-7\_15}.

\bibitemdeclare{inproceedings}{COSTCompetition2011}
\bibitem{COSTCompetition2011}
\bibinfo{author}{Thorsten \surnamestart Bormer\surnameend},
  \bibinfo{author}{Marc \surnamestart Brockschmidt\surnameend},
  \bibinfo{author}{Dino \surnamestart Distefano\surnameend},
  \bibinfo{author}{Gidon \surnamestart Ernst\surnameend},
  \bibinfo{author}{Jean-Christophe \surnamestart Filli{\^a}tre\surnameend},
  \bibinfo{author}{Radu \surnamestart Grigore\surnameend},
  \bibinfo{author}{Marieke \surnamestart Huisman\surnameend},
  \bibinfo{author}{Vladimir \surnamestart Klebanov\surnameend},
  \bibinfo{author}{Claude \surnamestart March{\'e}\surnameend},
  \bibinfo{author}{Rosemary \surnamestart Monahan\surnameend},
  \bibinfo{author}{Wojciech \surnamestart Mostowski\surnameend},
  \bibinfo{author}{Nadia \surnamestart Polikarpova\surnameend},
  \bibinfo{author}{Christoph \surnamestart Scheben\surnameend},
  \bibinfo{author}{Gerhard \surnamestart Schellhorn\surnameend},
  \bibinfo{author}{Bogdan \surnamestart Tofan\surnameend},
  \bibinfo{author}{Julian \surnamestart Tschannen\surnameend} \&
  \bibinfo{author}{Mattias \surnamestart Ulbrich\surnameend}
  (\bibinfo{year}{2011}): \emph{\bibinfo{title}{The COST IC0701 Verification
  Competition 2011}}.
\newblock In: {\sl \bibinfo{booktitle}{FoVeOOS'11}}, pp.
  \bibinfo{pages}{3--21}, \doi{10.1007/978-3-642-31762-0\_2}.

\bibitemdeclare{inproceedings}{Cohen:TPHOLs2009-23}
\bibitem{Cohen:TPHOLs2009-23}
\bibinfo{author}{Ernie \surnamestart Cohen\surnameend}, \bibinfo{author}{Markus
  \surnamestart Dahlweid\surnameend}, \bibinfo{author}{Mark \surnamestart
  Hillebrand\surnameend}, \bibinfo{author}{Dirk \surnamestart
  Leinenbach\surnameend}, \bibinfo{author}{Micha{\l} \surnamestart
  Moskal\surnameend}, \bibinfo{author}{Thomas \surnamestart Santen\surnameend},
  \bibinfo{author}{Wolfram \surnamestart Schulte\surnameend} \&
  \bibinfo{author}{Stephan \surnamestart Tobies\surnameend}
  (\bibinfo{year}{2009}): \emph{\bibinfo{title}{{VCC}: {A} Practical System for
  Verifying Concurrent {C}}}.
\newblock In: {\sl \bibinfo{booktitle}{Theorem Proving in Higher Order Logics
  (TPHOLs)}}, {\sl \bibinfo{series}{Lecture Notes in Computer Science}}
  \bibinfo{volume}{5674}, \bibinfo{publisher}{Springer}, pp.
  \bibinfo{pages}{23--42}, \doi{10.1007/978-3-642-03359-9\_2}.

\bibitemdeclare{techreport}{DeLine-Leino05}
\bibitem{DeLine-Leino05}
\bibinfo{author}{Rob \surnamestart DeLine\surnameend} \&
  \bibinfo{author}{K.~Rustan~M. \surnamestart Leino\surnameend}
  (\bibinfo{year}{2005}): \emph{\bibinfo{title}{{BoogiePL}: A Typed Procedural
  Language for Checking Object-oriented Programs}}.
\newblock \bibinfo{type}{Technical Report} \bibinfo{number}{MSR-TR-2005-70},
  \bibinfo{institution}{Microsoft Research}.
\newblock
  \urlprefix\url{ftp://ftp.research.microsoft.com/pub/tr/TR-2005-70.pdf}.

\bibitemdeclare{incollection}{FilliatreM04}
\bibitem{FilliatreM04}
\bibinfo{author}{Jean-Christophe \surnamestart Filli\^{a}tre\surnameend} \&
  \bibinfo{author}{Claude \surnamestart March\'{e}\surnameend}
  (\bibinfo{year}{2004}): \emph{\bibinfo{title}{Multi-prover Verification of
  {C} Programs}}.
\newblock In: {\sl \bibinfo{booktitle}{Formal Methods and Software
  Engineering}}, \bibinfo{series}{LNCS 3308}, \bibinfo{publisher}{Springer},
  pp. \bibinfo{pages}{15--29}, \doi{10.1007/978-3-540-30482-1\_10}.

\bibitemdeclare{inproceedings}{Goldstein2007}
\bibitem{Goldstein2007}
\bibinfo{author}{M.~\surnamestart Goldstein\surnameend}, \bibinfo{author}{Y.A.
  \surnamestart Feldman\surnameend} \& \bibinfo{author}{S.~\surnamestart
  Tyszberowicz\surnameend} (\bibinfo{year}{2006}):
  \emph{\bibinfo{title}{Refactoring with contracts}}.
\newblock In: {\sl \bibinfo{booktitle}{Agile Conference, 2006}}, pp.
  \bibinfo{pages}{10 pp. --64}, \doi{10.1109/AGILE.2006.44}.

\bibitemdeclare{mastersthesis}{GrebingDA2012}
\bibitem{GrebingDA2012}
\bibinfo{author}{Sarah \surnamestart Grebing\surnameend}
  (\bibinfo{year}{2012}): \emph{\bibinfo{title}{Evaluating and Improving the
  Usability of Interactive Verification Systems}}.
\newblock \bibinfo{type}{Diploma thesis}, \bibinfo{school}{Universit\"{a}t
  Koblenz-Landau}.

\bibitemdeclare{mastersthesis}{Hull2010}
\bibitem{Hull2010}
\bibinfo{author}{Ian \surnamestart Hull\surnameend} (\bibinfo{year}{2010}):
  \emph{\bibinfo{title}{{Automated Refactoring of Java Contracts}}}.
\newblock Master's thesis, \bibinfo{school}{University College Dublin}.

\bibitemdeclare{inproceedings}{Kaiser2007evolution}
\bibitem{Kaiser2007evolution}
\bibinfo{author}{Robert \surnamestart Kaiser\surnameend} \&
  \bibinfo{author}{Stephan \surnamestart Wagner\surnameend}
  (\bibinfo{year}{2007}): \emph{\bibinfo{title}{Evolution of the {PikeOS}
  Microkernel}}.
\newblock In \bibinfo{editor}{Ihor \surnamestart Kuz\surnameend} \&
  \bibinfo{editor}{Stefan~M \surnamestart Petters\surnameend}, editors: {\sl
  \bibinfo{booktitle}{MIKES: 1st International Workshop on Microkernels for
  Embedded Systems}}.

\bibitemdeclare{article}{Klein_EHACDEEKNSTW_10}
\bibitem{Klein_EHACDEEKNSTW_10}
\bibinfo{author}{Gerwin \surnamestart Klein\surnameend}, \bibinfo{author}{June
  \surnamestart Andronick\surnameend}, \bibinfo{author}{Kevin \surnamestart
  Elphinstone\surnameend}, \bibinfo{author}{Gernot \surnamestart
  Heiser\surnameend}, \bibinfo{author}{David \surnamestart Cock\surnameend},
  \bibinfo{author}{Philip \surnamestart Derrin\surnameend},
  \bibinfo{author}{Dhammika \surnamestart Elkaduwe\surnameend},
  \bibinfo{author}{Kai \surnamestart Engelhardt\surnameend},
  \bibinfo{author}{Rafal \surnamestart Kolanski\surnameend},
  \bibinfo{author}{Michael \surnamestart Norrish\surnameend},
  \bibinfo{author}{Thomas \surnamestart Sewell\surnameend},
  \bibinfo{author}{Harvey \surnamestart Tuch\surnameend} \&
  \bibinfo{author}{Simon \surnamestart Winwood\surnameend}
  (\bibinfo{year}{2010}): \emph{\bibinfo{title}{{seL4}: Formal Verification of
  an Operating System Kernel}}.
\newblock {\sl \bibinfo{journal}{Communications of the ACM}}
  \bibinfo{volume}{53}(\bibinfo{number}{6}), pp. \bibinfo{pages}{107--115},
  \doi{10.1145/1743546.1743574}.

\bibitemdeclare{incollection}{springerlink:10.1007/978-3-642-05089-3_51}
\bibitem{springerlink:10.1007/978-3-642-05089-3_51}
\bibinfo{author}{Dirk \surnamestart Leinenbach\surnameend} \&
  \bibinfo{author}{Thomas \surnamestart Santen\surnameend}
  (\bibinfo{year}{2009}): \emph{\bibinfo{title}{Verifying the Microsoft Hyper-V
  Hypervisor with VCC}}.
\newblock In \bibinfo{editor}{Ana \surnamestart Cavalcanti\surnameend} \&
  \bibinfo{editor}{Dennis \surnamestart Dams\surnameend}, editors: {\sl
  \bibinfo{booktitle}{FM 2009: Formal Methods}}, {\sl \bibinfo{series}{Lecture
  Notes in Computer Science}} \bibinfo{volume}{5850},
  \bibinfo{publisher}{Springer Berlin / Heidelberg}, pp.
  \bibinfo{pages}{806--809}, \doi{10.1007/978-3-642-05089-3\_51}.

\bibitemdeclare{inproceedings}{Maus08vx86}
\bibitem{Maus08vx86}
\bibinfo{author}{Stefan \surnamestart Maus\surnameend},
  \bibinfo{author}{Micha{\l} \surnamestart Moskal\surnameend} \&
  \bibinfo{author}{Wolfram \surnamestart Schulte\surnameend}
  (\bibinfo{year}{2008}): \emph{\bibinfo{title}{Vx86: x86 assembler simulated
  in C powered by automated theorem proving}}.
\newblock In: {\sl \bibinfo{booktitle}{12TH International Conference on
  Algebraic Methodology and Technology (AMAST 2008)}}, {\sl
  \bibinfo{series}{LNCS}} \bibinfo{volume}{5140},
  \doi{10.1007/978-3-540-79980-1\_22}.

\bibitemdeclare{inproceedings}{moura08z3}
\bibitem{moura08z3}
\bibinfo{author}{Leonardo \surnamestart de~Moura\surnameend} \&
  \bibinfo{author}{Nikolaj \surnamestart Bj{\o}rner\surnameend}
  (\bibinfo{year}{2008}): \emph{\bibinfo{title}{{Z3}: {A}n Efficient {SMT}
  Solver}}.
\newblock In: {\sl \bibinfo{booktitle}{Proc., 14th Int.\ Conf.\ on Tools and
  Algorithms for the Construction and Analysis of Systems, Budapest, Hungary}},
  \bibinfo{series}{LNCS 4963}, \bibinfo{publisher}{Springer}, pp.
  \bibinfo{pages}{337--340}, \doi{10.1007/978-3-540-78800-3\_24}.

\bibitemdeclare{book}{DO-178B}
\bibitem{DO-178B}
\bibinfo{author}{\surnamestart {RTCA SC-167 / EUROCAE WG-12}\surnameend}
  (\bibinfo{year}{1992}): \emph{\bibinfo{title}{DO-178B: Software
  Considerations in Airborne Systems and Equipment Certification}}.
\newblock \bibinfo{publisher}{Radio Technical Commission for Aeronautics
  (RTCA), Inc., 1828 L St. NW., Suite 805, Washington, D.C. 20036}.

\bibitemdeclare{phdthesis}{Sh12}
\bibitem{Sh12}
\bibinfo{author}{Andrey \surnamestart Shadrin\surnameend}
  (\bibinfo{year}{2012}): \emph{\bibinfo{title}{Mixed Low- and High Level
  Programming Language Semantics and Automated Verification of a Small
  Hypervisor}}.
\newblock Ph.D. thesis, \bibinfo{school}{Saarland University, Saarbrücken}.
\newblock
  \urlprefix\url{http://www-wjp.cs.uni-saarland.de/publikationen/Sh12.pdf}.

\bibitemdeclare{inproceedings}{SFM2010aprecisememorymodel}
\bibitem{SFM2010aprecisememorymodel}
\bibinfo{author}{Carsten \surnamestart Sinz\surnameend},
  \bibinfo{author}{Stephan \surnamestart Falke\surnameend} \&
  \bibinfo{author}{Florian \surnamestart Merz\surnameend}
  (\bibinfo{year}{2010}): \emph{\bibinfo{title}{A Precise Memory Model for
  Low-Level Bounded Model Checking}}.
\newblock In: {\sl \bibinfo{booktitle}{SSV'10}}.

\bibitemdeclare{misc}{Tobies2011}
\bibitem{Tobies2011}
\bibinfo{author}{Stephan \surnamestart Tobies\surnameend}:
  \emph{\bibinfo{title}{The Hyper-V Verification Experiment}}.
\newblock \bibinfo{note}{Presentation slides available at
  \url{http://research.microsoft.com/en-us/um/redmond/events/ss2011/slides/fri%
day/stephan_tobies.pdf}}.

\end{thebibliography}

\end{document}